# Warming early Mars with $CO_2$ and $H_2$

Ramses M. Ramirez[i,iii,v], Ravi Kopparapu[i,iii,v], Michael E. Zugger[ii,v], Tyler D. Robinson[iv,v], Richard Freedman[vi], and James F. Kasting[i,iii,v]

[i] Department of Geosciences, [ii] Applied Research Laboratory, [iii] Penn State Astrobiology Research Center, Penn State University University Park, PA 16802
[iv] Astronomy Department. University of Washington, Box 351580, Seattle, WA 98195
[v] NASA Astrobiology Institute Virtual Planetary Laboratory
[vi] SETI Institute, Mountain View, CA 94043/NASA Ames Research Center, Moffett Field, CA, 94035

## Abstract

**The presence of valleys on ancient terrains of Mars suggest that liquid water flowed on the martian surface 3.8 billion years ago or before. The above-freezing temperatures required to explain valley formation could have been transient, in response to frequent large meteorite impacts on early Mars, or they could have been caused by long-lived greenhouse warming. Climate models that consider only the greenhouse gases carbon dioxide and water vapor have been unable to recreate warm surface conditions, given the lower solar luminosity at that time. Here we use a one-dimensional climate model to demonstrate that an atmosphere containing 1.3-4 bar of $CO_2$ and water vapor, along with 5-20 percent $H_2$, could have raised the mean surface temperature of early Mars above the freezing point of water. Vigorous volcanic outgassing from a highly reduced early martian mantle is expected to provide sufficient atmospheric $H_2$ and $CO_2$ - the latter from the photochemical oxidation of outgassed $CH_4$ and CO - to form a $CO_2$-$H_2$ greenhouse. Such a dense early martian atmosphere is consistent with independent estimates of surface pressure based on cratering data.**

The climate of early Mars has been a topic of debate for at least the last 30 years. Nearly all researchers agree that the martian valleys and valley networks were formed by running water[1]. Debate has persisted as to how warm the surface must have been to form these features and how long this warmth must have lasted[2]. The widely cited impact hypothesis[3-5] suggests that large impacts occurring during the Heavy Bombardment Period could have heated the surface for brief intervals and that the valleys were formed by water that rained out following these events. Other authors[6] have argued that the martian climate was warmed by the greenhouse effect of a dense $CO_2$-$H_2O$ atmosphere, perhaps supplemented with $SO_2$[7]. But the $SO_2$ warming mechanism has difficulties because of photochemical production of sulfate aerosols, which act to cool the climate[8], and the calculation in ref. 7 is no longer believed because of errors in the $CO_2$ absorption coefficients[9]. Furthermore, all recent one-dimensional $CO_2$-$H_2O$ climate models[8,10,11] have been unable to produce above-freezing surface temperatures because of a combination of two factors: 1) $CO_2$ condensation, which reduces the tropospheric lapse rate, thereby lowering the greenhouse effect, and 2) Rayleigh scattering, which causes the planet's albedo to become high as the surface pressure becomes large[10]. Forget and Pierrehumbert[12] were able to produce mean surface temperatures above the freezing point of water by including explicit $CO_2$ ice clouds with 100 percent cloud cover. However, 3-D climate models predict much smaller fractional $CO_2$ cloud cover and greatly reduced surface warming.[13,14]

An updated 1-D climate calculation illustrates the basic problem (see Fig. 1). When solar luminosity is >80 percent of today's value, increased $CO_2$ partial pressure is capable of bringing Mars' mean surface temperature above 273 K. But for solar luminosities ≤80 percent of today, corresponding to time periods prior to ~2.8 Gyr ago[15], no amount of $CO_2$ can produce a warm surface. Instead, a dense $CO_2$ atmosphere would simply condense out globally in a 1-D climate model or at the poles in a more realistic 3-D climate model[14]. The perceived difficulty in producing a stable, warm climate on early Mars has bolstered support for the impact hypothesis, along with other "cold early Mars" theories.

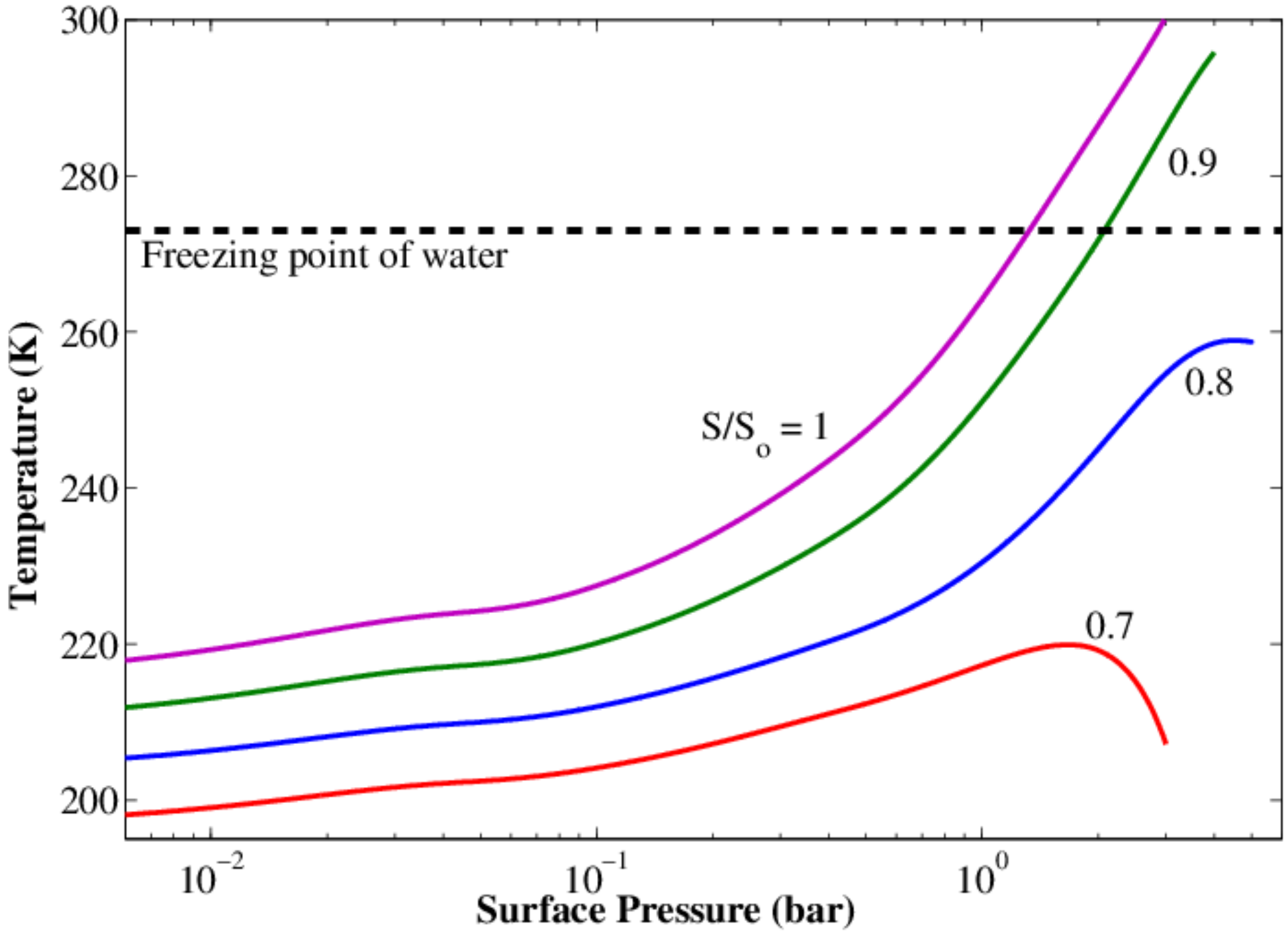

**Fig. 1:** Mean surface temperature as a function of surface pressure for a fully saturated (95% $CO_2$, 5% $N_2$) early Mars atmosphere at different solar insolation levels. The assumed surface albedo is 0.216. These results agree to within a few degrees with those of ref. (8), largely because increased absorption in the far wings of $CO_2$ and $H_2O$ lines (see Supp. Info) compensated for the loss of absorption arising from the updated $CO_2$ CIA parametrization (from ref. 11).

Since the impact hypothesis was proposed, other workers have pointed out that the valleys are more extensive than originally realized[16], and some have argued that the amount of time and surface runoff needed to form them was much larger than had been previously assumed[17,18]. Hoke et al.[18] performed detailed hydrologic modeling of several different valleys listed in their Table 3. The larger ones (*e.g.*, the 2°N, 34°E Naktong east valley) require episodic runoff rates of 0.5 cm/day, along with intermittent runoff averaging ~10 cm/yr for (3-4)×$10^7$ yr. These high flow rates are needed to lift eroded material off the river bed and keep it in suspension. According to Hoke et al., the estimated runoff total for this one valley is 3-4 million meters, or more than three orders of magnitude more than the amount of rainwater provided by the impact hypothesis. So, these authors envision an entirely different, and much wetter, scenario for martian valley formation. Here, we propose a mechanism for supplying these significantly larger rainfall amounts.

**Greenhouse warming by models that include $H_2$**

The utility of $H_2$ as a greenhouse gas for terrestrial planets was pointed out several years ago by Stevenson[19] and has been studied more quantitatively by recent authors.[20,21] Wordsworth and Pierrehumbert[21] showed that early Earth could have been kept warm by greenhouse warming from collision-induced absorption (CIA) caused by the interaction of $H_2$ molecules with $N_2$. These collisions excite the pure rotational levels of $H_2$ and, at the same time, allow it to absorb electromagnetic radiation to lift it from one rotational state to the next. At room temperature, the absorption spectrum of $H_2$ extends right through the 8-12 μm "window region", allowing it to be an effective greenhouse gas on either early Earth or Mars.[21,22] One difference between the two planets is that Mars has long been deficient in $N_2$[23], and so the main broadening gas there may instead have been $CO_2$. Collisional excitation of $H_2$ by $CO_2$ has not been studied, but there is no reason to suppose that it would be any less efficient than excitation by $N_2$. Indeed, collisional broadening of permitted absorption lines is stronger for $CO_2$ (Supp. Info.). Below, we conservatively assume that $H_2$-$CO_2$ CIA is of the same strength as $H_2$-$N_2$ CIA, for which the excitation cross sections have been calculated theoretically[24].

We included $H_2$ CIA, along with other updates, in our existing 1-D radiative-convective climate model[25]. Details are provided in Supp. Info. We then performed a series of climate calculations for hypothetical martian paleoatmospheres containing various amounts of $CO_2$ and $H_2$ (Fig. 2). The assumed solar luminosity was 0.75 times present, which is appropriate for 3.8 Gyr ago[15]. When $H_2$ was absent, the mean surface temperature never exceeded 230 K, regardless of how much $CO_2$ was present. This temperature is similar to values found previously by our group: published maximum temperatures for this solar flux were ~225 K[10] and 231 K[8]. Wordsworth et al.[11] could reach only 217 K for a 1-bar $CO_2$ atmosphere, but their model did not include $H_2O$.

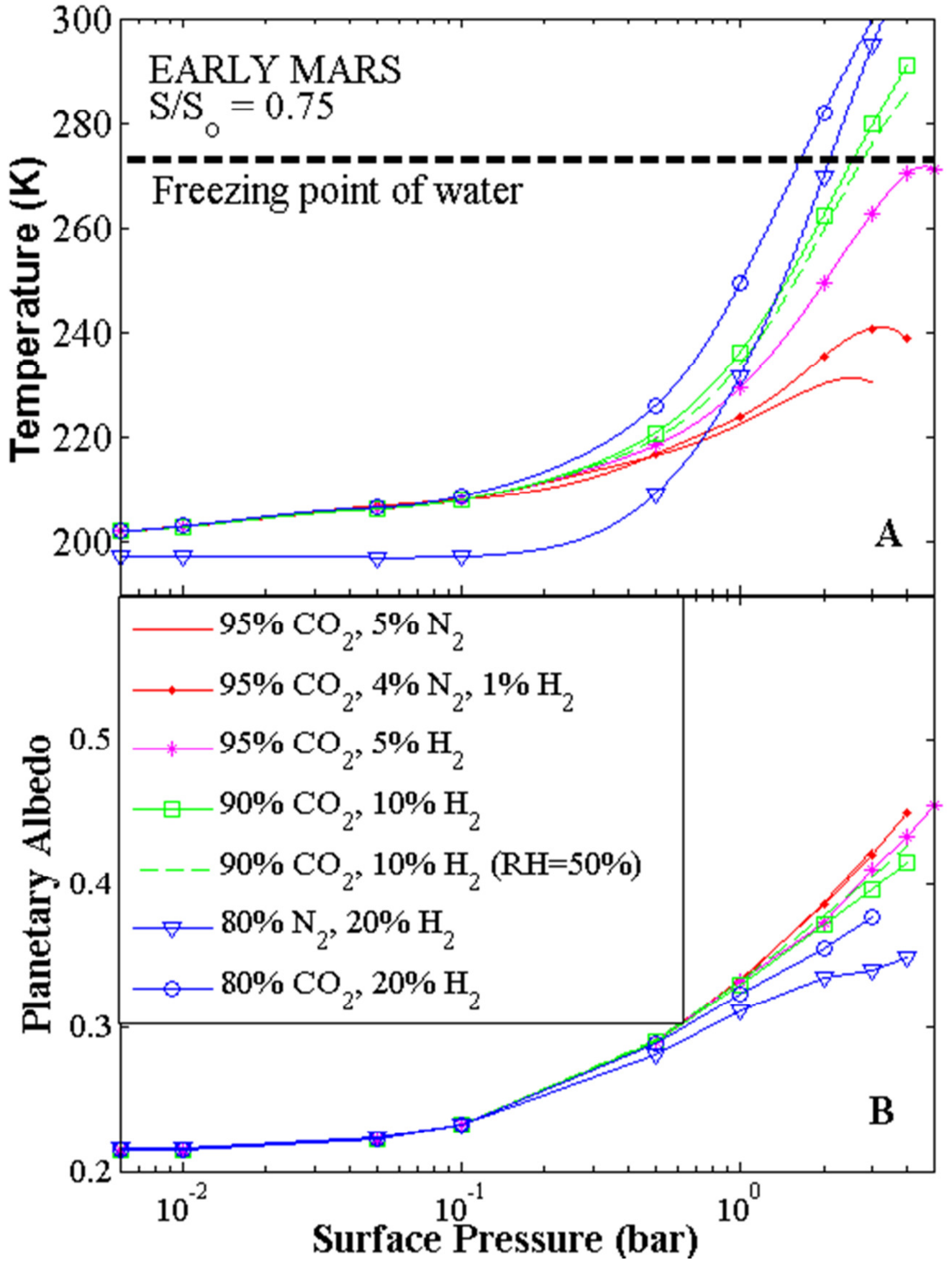

**Fig. 2:** Surface temperature (a) and planetary albedo (b) as a function of surface pressure for different atmospheric compositions. The assumed solar luminosity is 0.75 times present, appropriate for 3.8 Gyr ago. Any remaining gas not accounted for in the legend is considered to be $N_2$. Solid curves correspond to fully saturated atmospheres, while the model represented by the dashed curve assumes 50% tropospheric relative humidity. At 273 K and 10% $H_2$, a difference of only ~3-4 K separates the fully saturated and 50% relative humidity cases, indicating that the vast majority of the greenhouse warming is caused by $H_2$-$CO_2$ collision-induced absorption.

When 5 percent $H_2$ was included in our new climate model, the calculated maximum surface temperature nearly reached 273 K as $CO_2$ was increased, and it exceeded 273 K when the solar radiation calculation was done more accurately (Supp. Info.). The corresponding surface pressure for this simulation was ~4 bar (or ~3 bar in the more accurate calculation). Warming by $CO_2$ ice clouds[12,14], or $H_2O$ cirrus clouds[26], both of which were neglected in these calculations, would increase the mean surface temperature still further; thus, we take 5 percent as a reasonable minimum estimate for the $H_2$ concentration needed to produce a mean surface temperature above freezing. Higher $H_2$ concentrations allow the freezing point to be reached at lower surface pressures. For 10 percent $H_2$ this threshold was reached at ~2.5 bar, and for 20 percent $H_2$ it was reached at ~1.6 bar. This latter value is below the upper limit of 1.9 bar calculated recently by Kite et al.[27] based on the size distribution of secondary craters from impacts. Those calculations assume a weak soil target strength consistent with river alluvium; for a harder bedrock surface the upper limit on pressure approximately doubles[27]. $H_2$-$N_2$ collisions could also conceivably have warmed Mars to this temperature for $N_2$ partial pressures >2 bar, although $N_2$ is less likely to have been present in abundance, as already noted.

Finally, we acknowledge that the goal of reaching a mean surface temperature of 273 K may be somewhat artificial, as both temperatures and precipitation rates vary as a function of latitude and altitude. Wolf and Toon[28] used a 3-D climate model to show that half of Earth's early oceans could have remained unfrozen at a mean temperature of only 260 K, and the same could be true for early Mars[13,26]. So, more realistic climate simulations may show that the valleys could have formed at somewhat lower concentrations of $H_2$ and/or $CO_2$.

**Sources and sinks for $H_2$ and $CO_2$**

Both $H_2$ and $CO_2$ would have to have been abundant on early Mars for this greenhouse warming mechanism to have succeeded. We assume that early Mars, like early Earth, was volcanically active and would have released H, C, and S-bearing gases. We begin our analysis by considering outgassing of hydrogen on modern Earth. On Earth, the dominant outgassed form of H for subaerial volcanism is $H_2O$[29], which has an outgassing rate of $1.0\times10^{14}$ mol yr$^{-1}$ (ref. 30). The $H_2$:$H_2O$ ratio in subaerial volcanic gases can be estimated by considering thermodynamic equilibrium at typical outgassing conditions (1450 K, 5 bar pressure)[29] for the following reaction

$$2\ H_2O \xleftrightarrow{K_1} 2\ H_2 + O_2 \qquad (1)$$

The $H_2$:$H_2O$ ratio, $R$, is given by the relation

$$\frac{pH_2}{pH_2O} \equiv R = \left(\frac{K_1}{fO_2}\right)^{0.5} \qquad (2)$$

Here, $pH_2$ and $pH_2O$ are the partial pressures of $H_2$ and $H_2O$ in the released gas, and $fO_2$ is the oxygen fugacity of the system, which is set by the magma. The equilibrium constant, $K_1$, calculated from Gibbs free energies (see Supp. Info.), is $1.80\times10^{-12}$ atm. For the relatively oxidized Earth, the oxygen fugacity is near the QFM (quartz-fayalite-magnetite) synthetic buffer, for which $fO_2 \cong 10^{-8.5}$ at these P-T conditions[31], and so the predicted $H_2$:$H_2O$ ratio is 0.024. Multiplying by the outgassing rate for $H_2O$ gives a subaerial $H_2$ outgassing flux of $2.4\times10^{12}$ mol/yr, or $1\times10^{10}$ $H_2$ molec cm$^{-2}$ s$^{-1}$. (The actual $H_2$ outgassing rate scales as $R/(1+R)$, as can be demonstrated from mass balance, but this ratio reduces to $R$ when $R$ <<1.) This value is essentially identical with the estimate of total $H_2$ outgassing by Holland[32]. About half of Holland's total $H_2$ flux comes from submarine $H_2S$, which is readily photolyzed[8], and which can be converted to $H_2$ equivalents by the reaction: $H_2S + 2\ H_2O \rightarrow SO_2 + 3\ H_2$. (Further discussion of the atmospheric redox budget is given in Supp. Info.) Our calculated subaerial $H_2$ flux may thus be too high by a factor of 2, but by making this error we implicitly account for the contribution from $H_2S$. Scaling the outgassing rate in this manner might actually underestimate the $H_2S$ flux, considering that Mars' mantle appears to be exceptionally rich in sulfur[33]. The $H_2$ outgassing rate for modern Earth is uncertain by at least a factor of 5 (ref. 32), so trying to refine our estimate further would have limited benefit.

How might these outgassing fluxes scale for early Mars? It is difficult to be certain, partly because we do not know whether Mars ever experienced plate tectonics. We can make some crude analogies, however. The geothermal heat flux on Mars during the Noachian era is thought to have been similar to that of modern Earth[34], so it is plausible to assume that the total ($H_2O$+$H_2$) outgassing rate per unit area was the same on the two planets. The $H_2$ mole fraction should have been higher on Mars, though, because Mars' mantle is thought to be more reduced. Martian meteorites (SNCs) have $fO_2$'s ranging from QFM down to IW (iron-wűstite) or below[35,36]. (The IW $O_2$ buffer is about 4 log units below QFM.) The shergottites, which are the most primitive SNCs petrologically, and hence most like the mantle, have $fO_2$ values of ~IW+1[36]. And ALH84001, the oldest martian meteorite, is at IW−1, suggesting that the early martian mantle may have been even more reduced[36]. This prediction is in conflict with a recent analysis based on S and Ni abundances and the Mn/Fe ratio of Gusev crater rocks, which suggest that the upper part of Mars' early mantle was more oxidized in the distant past[37]. But these data might also be explained by later alteration[37], so the bulk of the evidence still favors a highly reduced early mantle.

Assume for the moment that Mars' mantle $fO_2$ was near IW+1, roughly 3 log units below the terrestrial value. Eq. (2) then predicts that the $H_2$:$H_2O$ ratio, $R$, in the released gas should increase by a factor of 32. The rate of $H_2$ outgassing should increase by a smaller amount ($\propto R/(1+R)$), assuming that the total outgassing rate of ($H_2$+$H_2O$) remains constant. . The three-log-unit decrease in mantle $fO_2$ compared to Earth should increase $H_2$ outgassing by a factor of ~20, so the expected $H_2$ outgassing rate on early Mars should be of the order of ($1\times10^{10}$ cm$^{-2}$s$^{-1}$)×20 = $2\times10^{11}$ cm$^{-2}$ s$^{-1}$. This direct flux of $H_2$ could have been augmented by ~50% by an indirect flux of hydrogen from $CH_4$ (see below and Supp. Info.), but we neglect that contribution here. This input of hydrogen should have been balanced by escape of hydrogen to space. If hydrogen escaped at the diffusion limit[38], the escape rate in these same units would be given by

$$\Phi_l(H_2) = \frac{b_i}{H_a} \cdot \frac{f_T(H_2)}{1+f_T(H_2)} \cong \frac{b_i f_T(H_2)}{H_a} \qquad (2)$$

Here, $b_i$ is a weighted molecular diffusion coefficient for H and $H_2$ in air (or $CO_2$), $H_a$ $(=kT/mg)$ is the pressure scale height, and $f_T(H_2)$ is the total hydrogen volume mixing ratio: $f_T(H_2) = 0.5f(H) + f(H_2) + f(H_2O) + ...$ If $H_2$ is the dominant H-bearing species, then $f_T(H_2) \cong f(H_2)$. All quantities are evaluated at the homopause, near 130 km altitude on present Mars, where molecular diffusion begins to exceed mixing by turbulent eddies. Counter-intuitively, Mars' lower gravity *decreases* the hydrogen escape rate compared to Earth because the scale height, $H_a$, appears in the denominator of eq. (2). For a homopause temperature of ~160 K, $b_i/H_a \cong 1.6\times10^{13}$ cm$^{-2}$ s$^{-1}$. Inserting this value into eq. (2) and equating $H_2$ escape with $H_2$ outgassing yields $f(H_2) \cong 0.013$. This is a factor of 4 less than the minimum $H_2$ mixing ratio of 0.05 needed to produce Earth-like surface temperatures on early Mars, according to Fig. 2. At a mantle $fO_2$ of IW−1, the $H_2$ outgassing flux would be another factor of 2 higher, making it $4\times10^{11}$ $cm^{-2}$ $s^{-1}$ and yielding an atmospheric $H_2$ mixing ratio of ~0.025, about half the minimum $H_2$ concentration needed to produce a warm early Mars (Fig. 2).

These numbers are all very uncertain, though. Volcanic outgassing rates on early Mars could have been considerably higher than those on modern Earth, and/or hydrogen may have escaped at less than the diffusion-limited rate. Calculations with our own 1-D hydrodynamic escape model, which is based on that of ref. 39, suggest that the diffusion limit would have been achieved. However, such 1-D calculations may overestimate the actual escape rate. The escape geometry is at least 2-dimensional, as solar radiation impinges on only one side of the planet. A third dimension is required if magnetic fields are included and if the planet's magnetic axis is inclined relative to its orbital plane. For neutral, nearly isothermal atmospheres escaping from hot Jupiter exoplanets, geometry alone reduces the mass loss rate by nearly a factor of 4 (ref. 40). That would bring the predicted atmospheric $H_2$ mixing ratio in our model up to 10 percent if Mars' mantle was at IW−1, or 5 percent at IW+1. Accounting for partial ionization of the escaping gases and their interaction with a planetary magnetic field, which might have existed up until ~3.9-4.0 Ga[41], might slow the escape even further, so $H_2$ mixing ratios of 20 percent are not impossible. More complicated magnetohydrodynamic escape models would be needed to study this possibility.

Availability of $CO_2$ must also be considered. On Earth, the atmospheric $CO_2$ abundance is controlled over long time scales by the balance between production by volcanism and loss by weathering of silicate minerals followed by precipitation of carbonates. $CO_2$ is predicted to accumulate in the atmosphere when the climate is cold because of slower weathering[42], and this feedback process could have been important on early Mars, as well[6]. On Mars, however, the total carbon inventory is uncertain, and additional terms in the $CO_2$ budget must be considered.

Carbonate minerals are observed in martian dust and in the SNC meteorites[43,44]. If the carbonate content of the dust is representative of the uppermost 2 km of crust, then the crust could contain the equivalent of 5 bar of $CO_2$[44]. This inference is speculative, however, and the apparent lack of large carbonate outcrops in the martian subsurface[43] suggests that much of Mars' $CO_2$ was lost by escape to space. $CO_2$ can be lost by both thermal[45] and nonthermal[46] loss processes. Thermal escape, which might initially have been very fast, presents the greatest hurdle for maintaining a dense early martian atmosphere. Thermal (hydrodynamic) loss of both C and O is predicted when the solar extreme ultraviolet (EUV) flux is more than ~5 times the modern solar mean, which would likely have been the case during the first several hundred million years of Mars' history[47]. Modeling suggests that the best time to accumulate $CO_2$ would be 3.8-4.0 Gyr ago, when solar activity had declined but when volcanoes were still active[45].

A second concern is whether enough $CO_2$ would have been supplied from volcanoes to maintain a dense atmosphere. Phillips et al.[48] estimated that 1.5 bar of $CO_2$ would have been emitted by Tharsis alone. But, because of Mars' much lower mantle oxygen fugacity, some authors have suggested that most of Mars' carbon would have been retained in the mantle as graphite[35,36]. These studies estimate upper limits of ~1 bar on outgassed $CO_2$ if Mars' mantle $fO_2$ was near IW+1. Just recently, however, Wetzel et al.[49] have studied this system experimentally and have shown that carbon would be stored in reduced silicate melts as a mixture of iron carbonyl, $Fe(CO)_5$, and $CH_4$. Upon outgassing, iron carbonyl would dissociate to form CO. According to their estimates, initial solidification of a 50 km-thick crust would have released 1 bar of CO and 1.3 bar of $CH_4$. Once in the atmosphere, both of these gases would have been oxidized by the byproducts of water vapor photolysis[25], yielding ~5.2 bar of $CO_2$ as an initial atmospheric inventory. $CO_2$ would have been continuously removed by carbonate formation and by escape, so recycling and/or continued juvenile outgassing would have been needed to maintain its abundance. To complicate matters, CO might have been more abundant than $CO_2$ if the carbon was outgassed in reduced form[50]. Although CO is not an effective greenhouse gas, calculations with $N_2$-$H_2$ mixtures (Fig. 2a) show that the surface can still be warmed above the freezing point even if the accompanying gas is not radiatively active. $CH_4$ is an effective greenhouse gas on modern Earth; however, our calculations indicate that its contribution to Mars' greenhouse effect was negligible (see Supp. Info.).

In summary, the formation of valleys and valley networks on early Mars is most easily explained if the climate was warm for long time periods. An $H_2$-$CO_2$-$H_2O$ greenhouse is capable of sustaining mean surface temperatures > $0^oC$ at 3.8 Gyr ago, provided that $H_2$ and $CO_2$ were maintained at reasonably high concentrations (5 percent for $H_2$, >1.3 bar for $CO_2$) by volcanic outgassing. CO also works in combination with $H_2$ at slightly higher surface pressures. High concentrations of $H_2$ would have been promoted by outgassing from a strongly reduced martian mantle, perhaps augmented by geometric or magnetic limitations on the rate at which $H_2$ escaped to space. $CO_2$ could have been supplied either directly if the mantle was relatively oxidized or indirectly as CO and $CH_4$ if the mantle was highly reduced. More detailed exploration of Mars' surface, including searches for buried carbonates, could provide additional evidence to test this hypothesis.

**Methods Summary**

The climate calculations described here were performed with a 1-dimensional (horizontally averaged) global climate model. Details and references are provided in Supp. Info. In such a model, the planet is assumed to be flat, and the Sun is placed at a solar zenith angle of $60^0$ from the vertical. Because $\cos(60^0) = 0.5$, multiplying the resulting solar fluxes by another factor of 0.5 gives the planetary average mean solar flux of $S/4$, where $S$ is the solar constant at the planet's orbit. Some sensitivity tests were performed with a more accurate, Gaussian integration over solar zenith angles (see Supp. Info.). In both the solar and thermal-infrared parts of the radiation code, the diffuse flux was calculated using a 2-stream approximation in which the integration over the upward and downward hemispheres is accomplished by choosing a single average zenith angle. In such a model, the stratosphere is assumed to be in radiative equilibrium, that is, the net emitted infrared flux is equal to the net absorbed solar flux in each layer. In the troposphere (nearer to the planet's surface), convection was accounted for by using a moist convective adjustment, as follows: If the temperature lapse rate calculated from radiative equilibrium exceeded the moist adiabatic lapse, the lapse rate in that layer was adjusted back to the moist adiabat. Here, the term "moist" refers to saturation with either $H_2O$ or $CO_2$. Typically, $H_2O$ condenses within the lower part of the planet's troposphere, whereas $CO_2$ condenses in the upper troposphere. As described in the main text, the troposphere was assumed to be fully saturated with $H_2O$ in most of our simulations. This produces a slight overestimate for the surface temperature, but the error should be generally small (see Fig. 2). More importantly, the effect of clouds on the planetary albedo is not calculated accurately in our model; instead, the atmosphere is assumed to be cloud-free, and the surface albedo is adjusted to a value (0.216) that allows the model to reproduce Mars' current mean surface temperature, 218 K, given current solar insolation. This surface albedo is held fixed for all the calculations shown in Figs. 1 and 2. More complex, 3-D climate models are needed to treat clouds and relative humidity more realistically.

Requests for correspondence and additional materials should be addressed to Ramses Ramirez: ramsesny@gmail.com

Acknowledgements

This paper benefited from reviews by Brian Toon, Robin Wordsworth, and an anonymous reviewer. We are grateful to Lee Kump for the tip on iron carbonyl. We also thank Vincent Eymet for providing KSPECTRUM. Support for this work came from the NASA Exobiology Program and the NASA Astrobiology Institute.

Individual contributions: RR and RK generated $H_2O$ and $CO_2$ line-by-line cross-sections. RR generated $CH_4$ line-by-line cross sections with guidance from RF. RR performed most of the background research and climate model updates. RR and RK debugged the climate model. RR performed the computations and wrote most of Supp. Info. TR worked with RR in providing flux comparisons with SMART, and MZ performed numerical calculations of hydrodynamic escape rates. JK provided overall guidance and wrote much of the main text. All authors contributed to proof-reading and making comments on the paper.

# SUPPLEMENTARY INFORMATION

## 1. CLIMATE MODEL DESCRIPTION

We used a 1-D radiative-convective climate model[51,52,53] to study $CO_2$-$H_2$ greenhouse warming on early Mars. The current version of the model divides the atmosphere into 100 unevenly spaced layers in log pressure extending from the ground to a pressure of $3\times10^{-5}$ bar. Radiative equilibrium is assumed for each layer in the stratosphere. At lower altitudes, if the radiative lapse rate within a layer exceeds the moist adiabatic lapse rate, then a convective adjustment is performed. The model relaxes to a moist $H_2O$ adiabat at higher temperatures or to a moist $CO_2$ adiabat when it is cold enough for $CO_2$ to condense[54]. This defines a convective troposphere near the surface. Within this troposphere, water vapor is assumed to be fully saturated for 9 out of our 10 simulations, thereby maximizing the greenhouse effect. (See main text, Fig. 2.) For the 95% $CO_2$/5% $H_2$ case, a sensitivity run with 50 percent relative humidity was performed to determine how much difference this would make. $H_2O$ and $CO_2$ clouds are neglected in the model, but the effect of the former is accounted for by increasing the surface albedo, as done in previous climate simulations by our group[53,54]. It has been argued that this methodology tends to overestimate the greenhouse effect of dense early atmospheres[55]. By contrast, our neglect of $CO_2$ clouds may cause us to underestimate the greenhouse effect[56,57]. Realistically determining the effect of clouds would probably require a 3-D climate model, and even then the cloud problem remains difficult.

Incident solar radiation and outgoing thermal infrared radiation are both treated using a two-stream approximation[58]. Delta function scaling and the quadrature method are used at solar wavelengths. At those wavelengths, the model uses correlated-k coefficients to parameterize absorption by $CO_2$, $H_2O$ and $CH_4$ across 38 spectral intervals ranging from 0.2 to 4.5 microns. In the solar, we currently use 8-term $CO_2$ k-coefficients derived from HITRAN 2008, whereas our 8-term $H_2O$ coefficients utilize HITRAN 2008 at low pressures and HITEMP 2010 for pressures greater than or equal to 0.1 bar[59]. The near-IR $CH_4$ coefficients for wavelengths less than 1 μm are those derived from Karkoschka[60], whereas those longward of 1 μm were derived from self-broadened k-distributions[61]. These latter coefficients were interpolated between 188 K and 295 K. These $CO_2$, and $H_2O$ coefficients, and the thermal-IR ones described below, have replaced those used in our other recent studies of early Mars[57,62], and they produce excellent agreement in both the upwelling and downwelling thermal-IR fluxes when compared against both the well-tested SMART line-by-line model[63] and Fig. 2c of Wordsworth et al.[64] for the same dense $CO_2$ atmosphere used by the latter authors (see Fig. S1-S5).

In the thermal infrared, 8-term k-coefficients are used for $CO_2$ and $H_2O$[59] and new 4-term coefficients have been derived for $CH_4$, in 55 spectral intervals extending from 0 – 15,000 $cm^{-1}$. The coefficients for $CO_2$ and $H_2O$ were calculated using line width truncations of 500 $cm^{-1}$ and 25 $cm^{-1}$, respectively, and were computed over 8 temperatures (100, 150, 200, 250, 300, 350, 400, 600 K) and 8 pressures ($10^{-5}$ – 100 bar). A short truncation width was used for $H_2O$ because we overlaid this with a continuum, as described below. The $CH_4$ coefficients (truncated at 35 $cm^{-1}$) were computed at the same 8 pressures and the following 5 temperatures (100,200,300,400,600) K. Overlap between gases was computed by convolving the k-coefficients within each broadband spectral interval.

We use the 4.3 micron $CO_2$ chi factors of Perrin and Hartmann[65] as a proxy for $CO_2$ far wing absorption in the 15 micron region. For water, the BPS continuum of Paynter and Ramaswamy[66] is overlain over its entire range of validity (0 – 19,000$cm^{-1}$).

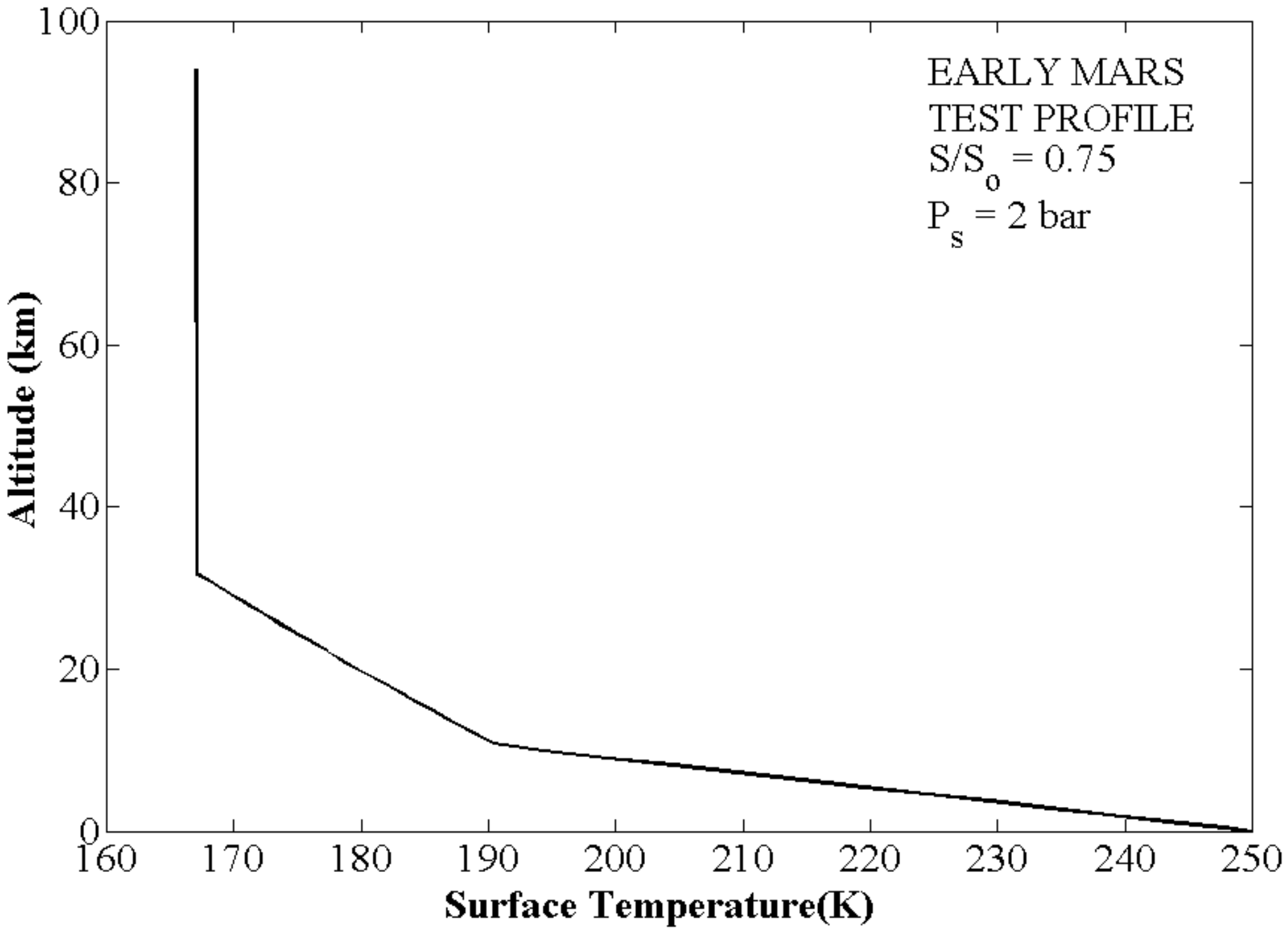

**Figure S1**: Temperature-pressure profile for a 2-bar 95% $CO_2$ 5 % $N_2$ dry atmosphere with early Mars insolation ($S/S_o$ = 0.75) used for flux comparisons in Fig. S2- S5. The surface temperature is 250 K and stratospheric temperature is fixed at 167K. This test profile is not an output of the nominal runs in the main text.

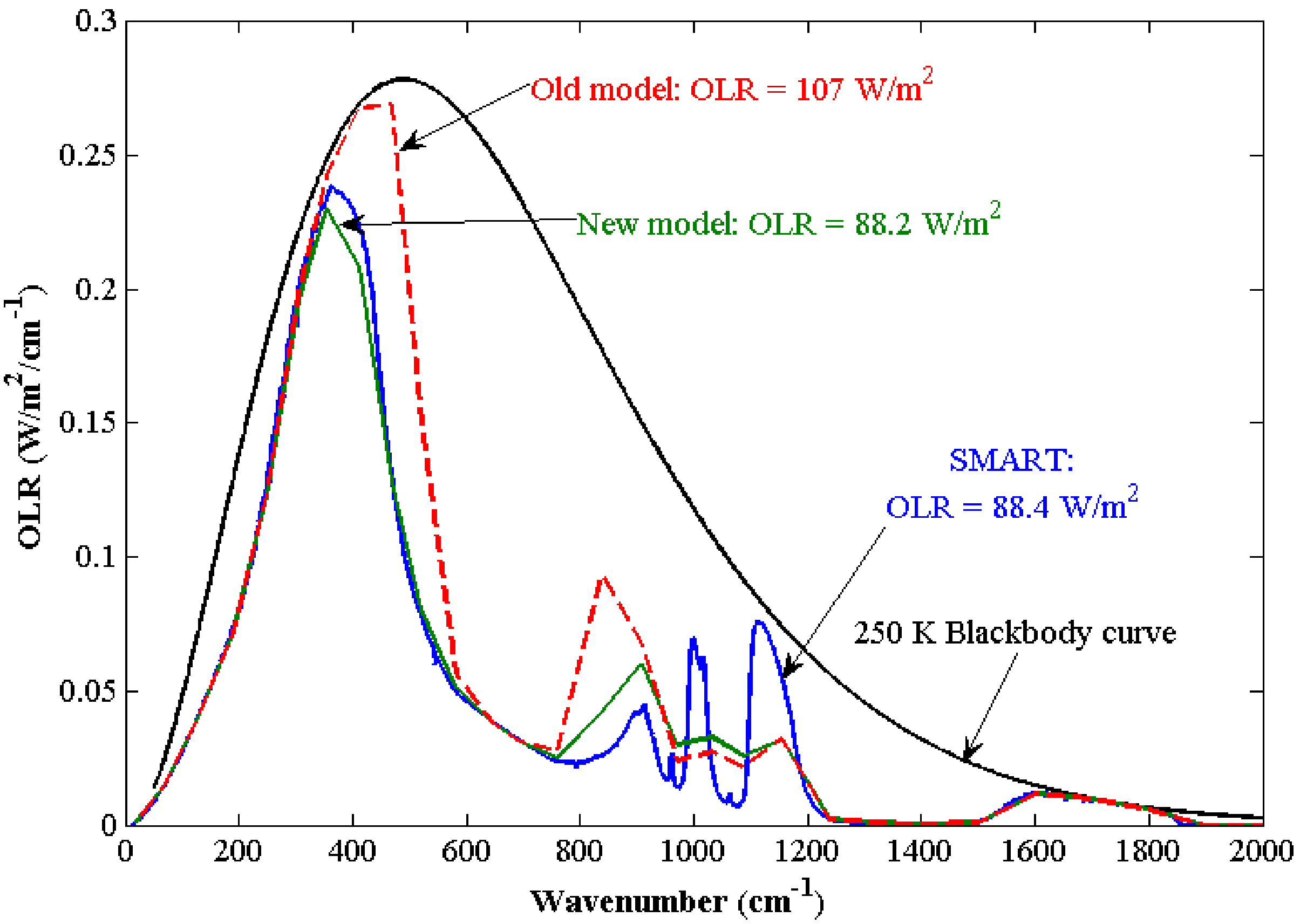

**Figure S2:** Emission spectra for the atmosphere in Fig. S1, comparing (a) the old model $CO_2$ coefficients against those of both (b) the new model (this paper), and (c) the line-by-line model SMART[63]. All three models use the GBB $CO_2$ CIA parameterization.[73,74,75] Our computed outgoing longwave radiative (OLR) flux (88.2 W/m$^2$) is in excellent agreement with Wordsworth et al.[64] Fig 2c and comes to within ~0.2% of SMART. Slight differences between our new model and SMART are attributable to spectral resolution differences and different assumptions about line profiles. Our model uses Voigt line profiles, whereas SMART utilizes van Vleck-Weisskopf and Rautian line profiles. Our old model underestimates $CO_2$ absorption by ~20 W/m$^2$ because of relatively short line truncation (~25 cm$^{-1}$), which neglects much of the far wing contribution at higher pressures[72]. Note that previous simulations by our group using the old model[52,53,62] had implemented the Kasting et al.[51] $CO_2$ CIA parameterization, which overestimated continuum absorption. Here, we decided to use the GBB $CO_2$ CIA parameterization for ease of comparison with all three cases.

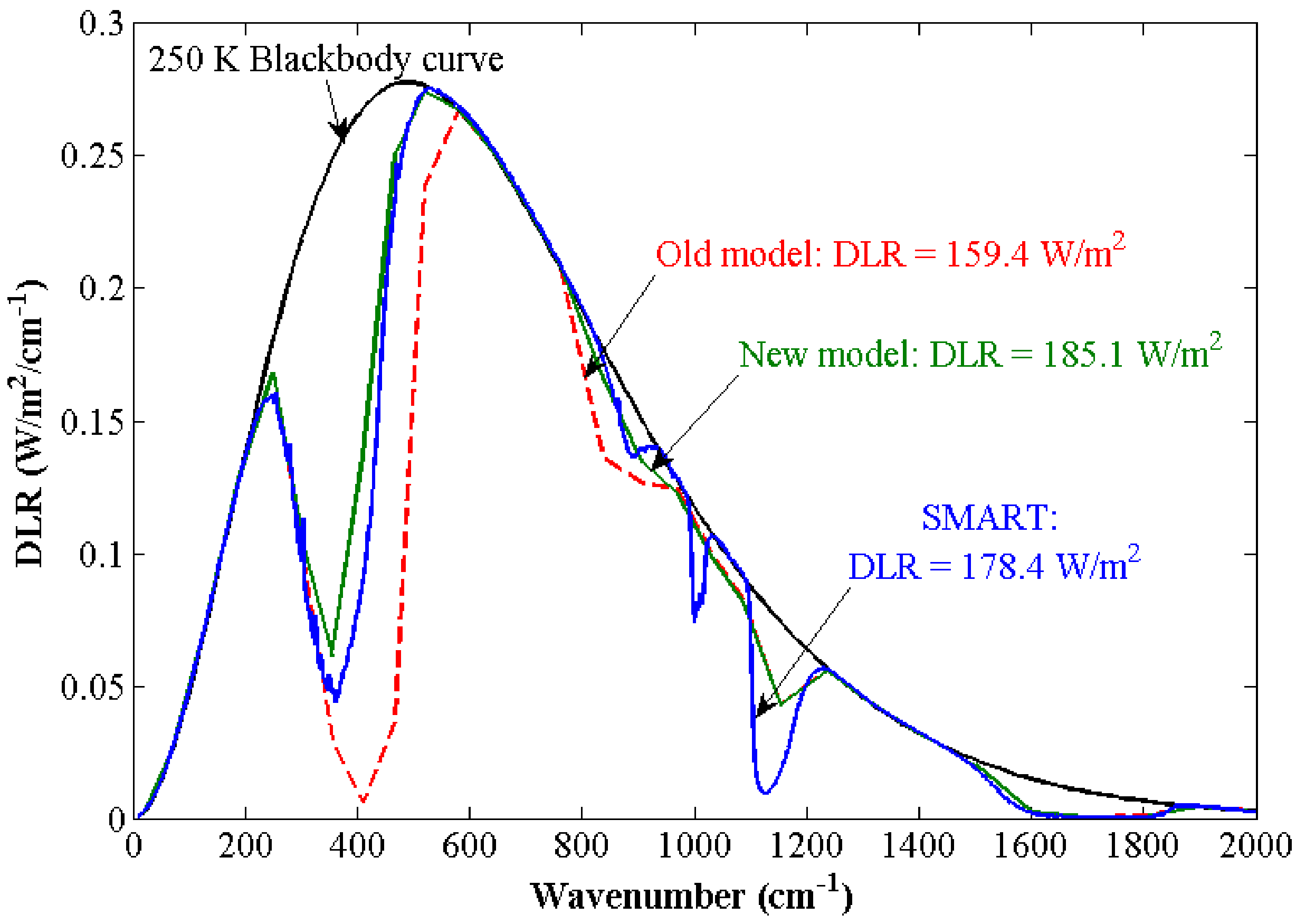

**Figure S3:** Surface downwelling longwave radiative (DLR) spectral flux comparisons for the atmosphere in Fig. S1. The longer $CO_2$ line shapes used by both our new model and by SMART explain the ~14% and 11% higher surface downwelling fluxes, respectively, as compared to our old model. Our slightly larger (~3.6%) surface flux as compared to SMART is likely due to the line profile differences discussed in the caption to Fig. S2.

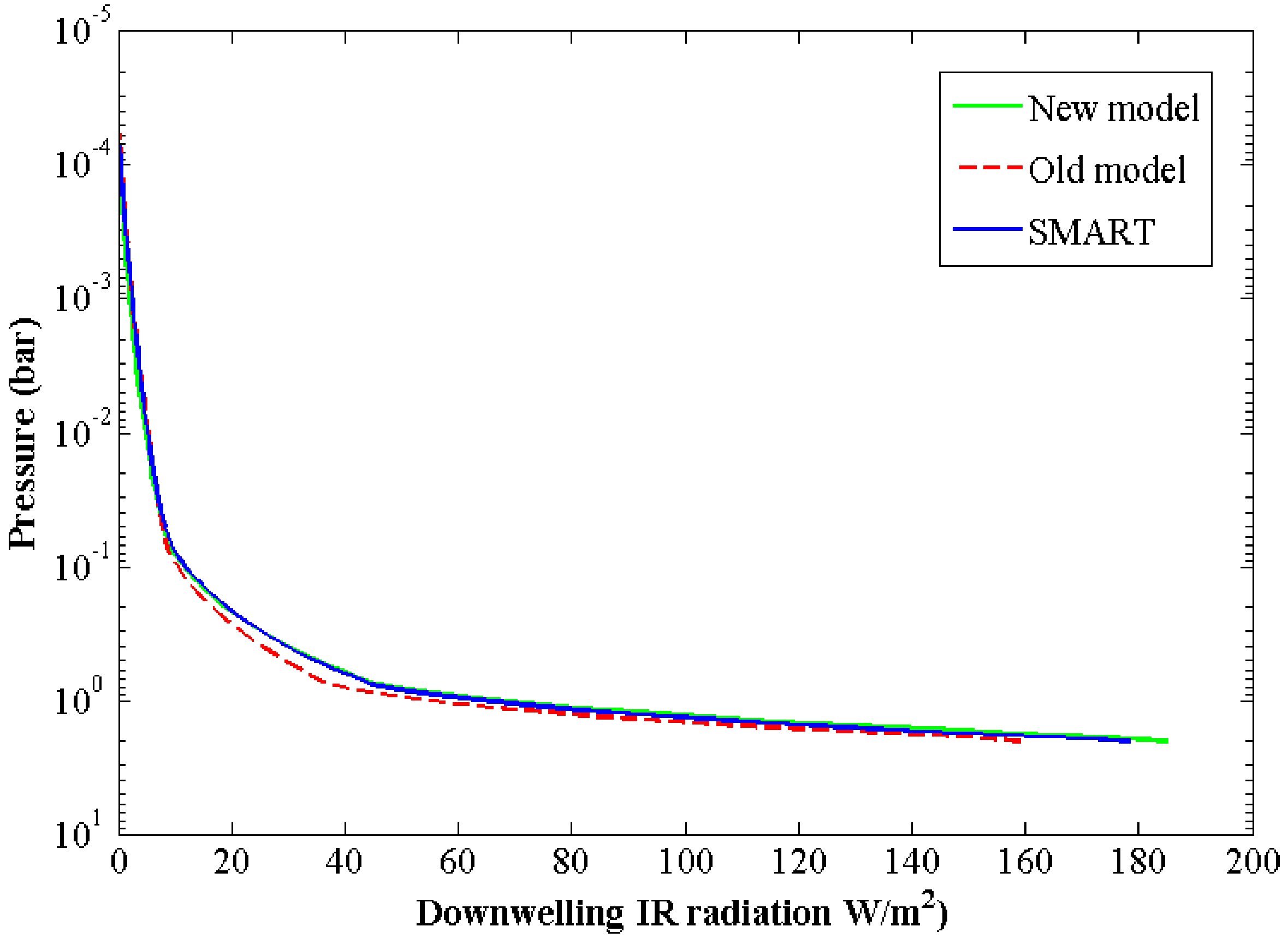

**Figure S4:** Downwelling longwave radiative (DLR) broadband flux comparisons for the atmosphere in Fig. S1. Fluxes calculated by SMART and by our new model agree well at most heights, although our model has a slightly stronger (~3.6%) greenhouse effect at the surface.

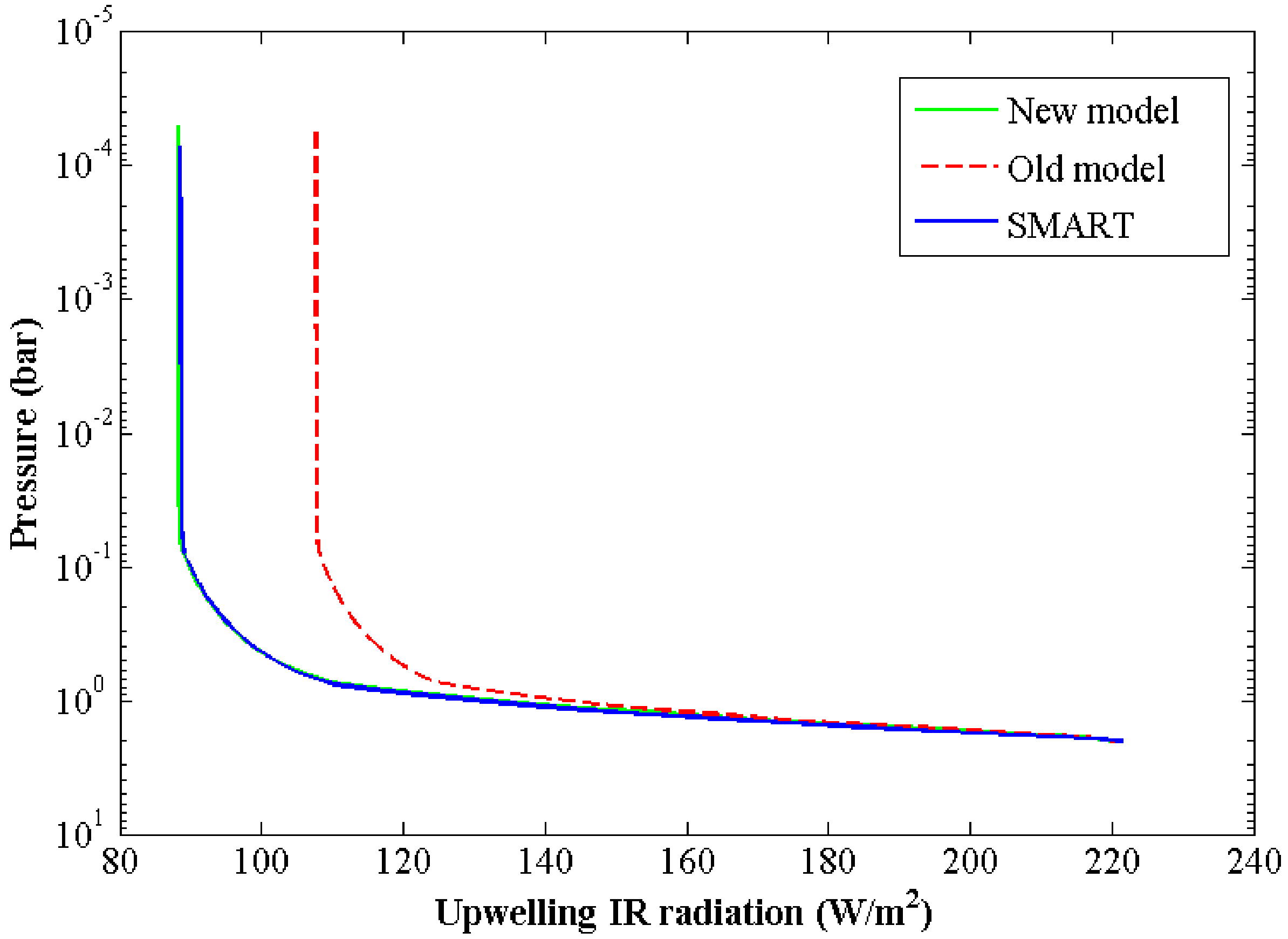

**Figure S5:** Upwelling longwave radiative broadband flux comparisons for the atmosphere in Fig. S1. Our new model is in excellent agreement with SMART at all heights, whereas the flux calculated by our old model is higher by as much as 19.4 $W/m^2$ in the stratosphere.

The following additional updates were made to the climate model:

1) We included collision-induced absorption (CIA) caused by the interaction of $H_2$ with $CO_2$. $CO_2$ is an anisotropic molecule that picks up a strong dipole moment when its bending mode is activated, necessitating extra quantum mechanical treatments that are at this stage considered experimental (L. Frommhold, personal communication). Therefore, we modeled $CO_2$-$H_2$ CIA using measured absorption coefficients for $N_2$-$H_2$ CIA[67]. We believe that this approach is conservative, as Burch et al.[68] determined that self-broadening of permitted $CO_2$ transitions was ~30% more effective than foreign-broadening by $N_2$. The relatively small moment of inertia for $H_2$, coupled with its widely-spaced rotational levels, results in absorption over a large swath of the thermal infrared, including the 200-600 $cm^{-1}$ and the 8-12 micron regions[69] (See Figs. S6 and S7).

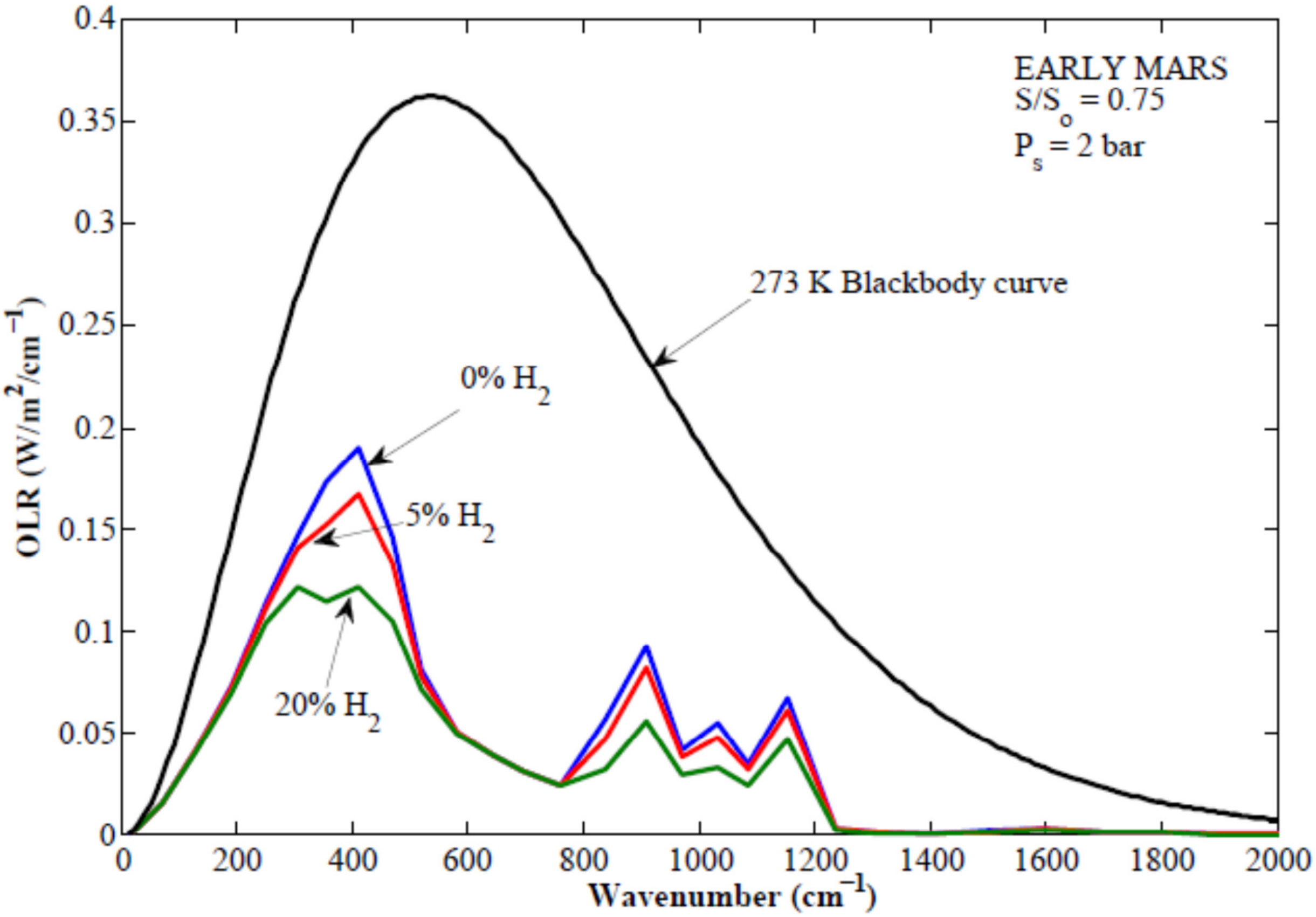

**Figure S6:** Emission spectrum for a 2-bar early Mars ($S/S_o = 0.75$) atmosphere containing 95% $CO_2$ and 5% $N_2$ (blue), 95% $CO_2$ and 5% $H_2$ (red), or 80% $CO_2$ and 20% $H_2$ (green). The surface temperature is 273K, and the stratospheric temperature is fixed at 167 K. Adding 5% and 20% $H_2$ reduces the outgoing infrared flux by ~6 W/m$^2$ and 22 W/m$^2$, respectively. Vertical infrared flux profiles for these cases are shown in Fig. S7.

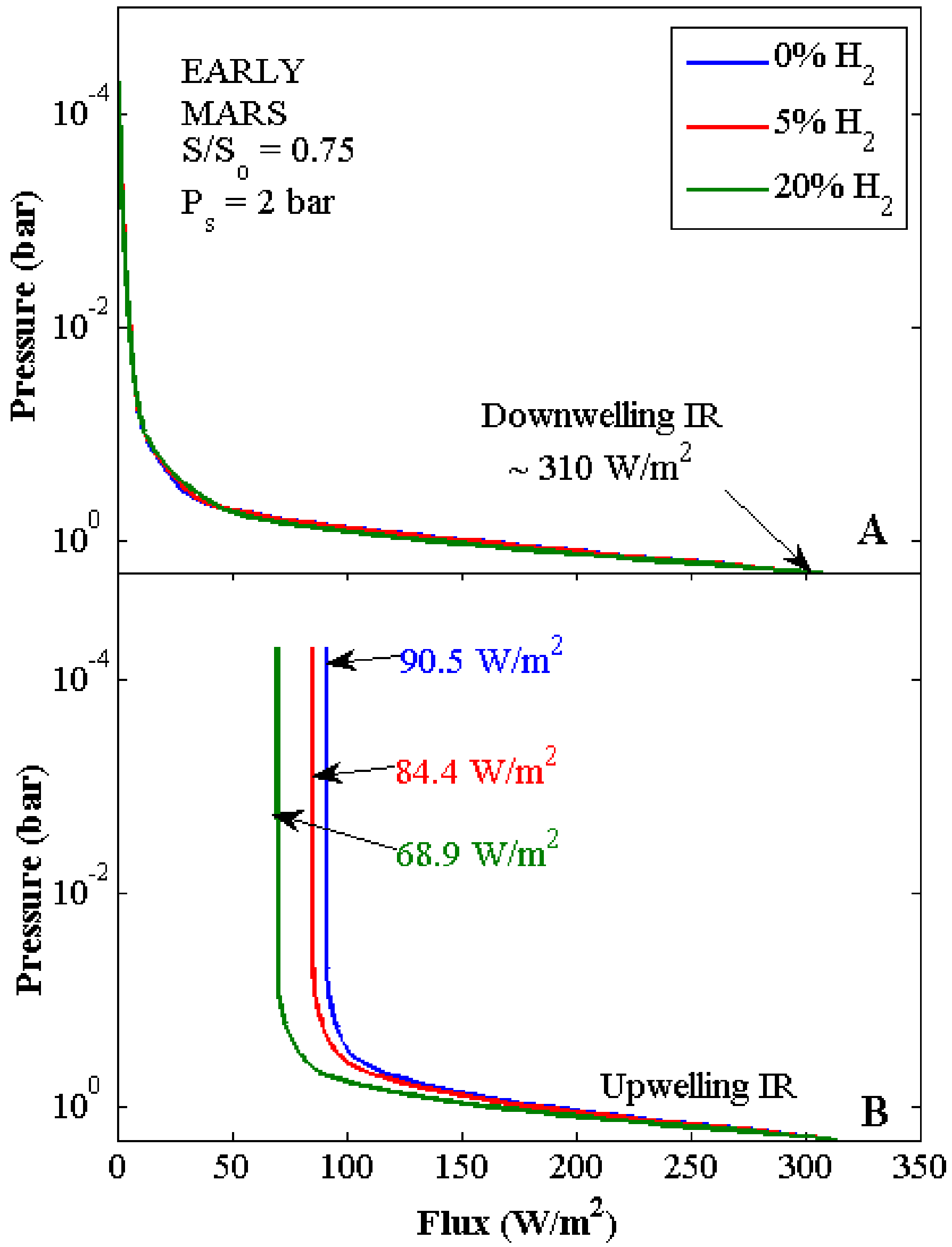

**Figure S7:** Infrared (IR) flux profiles for the 2-bar, 273 K surface temperature atmosphere shown in Figure S6. Although all scenarios yielded a similar downwelling IR flux, the 5% and 20% $H_2$ cases resulted in ~ 6 and 22 W/m$^2$ reductions in upwelling IR, respectively, greatly enhancing the greenhouse effect. Coupled with a strong $CO_2$ greenhouse effect, these radiative forcings from $H_2$ increase atmospheric temperatures above freezing.

Indeed, collision-induced absorption by $H_2$ broadened by other molecules such as Ar, $N_2$, or even itself, all produce absorption over the same spectral range[70]. Subsequently, the same would be true for $CO_2$-$H_2$. Although $CO_2$ is more effective at Rayleigh scattering than is $N_2$, our $CO_2$-$H_2$ cases still produce more surface warming than does our $N_2$-$H_2$ case (Fig. 2, main text). Thus, the increase in Rayleigh scattering as $CO_2$ increases is outstripped by increased absorption within the 15 micron band. This suggests that $CO_2$ would have to broaden $H_2$ considerably less effectively than $N_2$ does for our mechanism to fail, contrary to the available evidence[68]. Consequently, we use the $N_2$-$H_2$ CIA data of Borysow and Frommhold[67] as a lower bound for the calculation of $CO_2$-$H_2$ opacity ($\tau_l$). This lower bound parameterization is then:

$$\tau_l = H_i \frac{n}{n_o} W_{H_2} f_{CO_2} \tag{S1}$$

Here, $H_i$ is a constant (see below) with units of amagat$^{-2}$ $cm^{-1}$, $W_{H2}$ is the $H_2$ path length in atmosphere-centimeters, $n$ is number density, $n_o$ is Loschmidt's constant ($2.687\times10^{19}$ molecules/atm-$cm^3$), and $f_{H2}$ is the $H_2$ mixing ratio. The quantity $f_{CO2}$ represents the foreign broadening by $CO_2$. This lower bound estimate resulted in mean surface temperatures above 273 K at ~ 2.5 bar for 10% $H_2$ and ~1.5 bar for 20% $H_2$.

The $N_2$-$H_2$ collision-induced absorption coefficients, $H_i$, used in Eq. S1 are tabulated below (Table I). $N_2$-$H_2$ CIA coefficients, like other absorption coefficients in our model, are interpolated log-linearly (linear in $T$, logarithmic in the absorption coefficient) between the values shown in Table I.

2) We also incorporated self-broadening by $H_2$-$H_2$ pairs[71], using the following equation for the opacity, $\tau_p$:

$$\tau_p = H_i \frac{n}{n_o} W_{H_2} f_{H_2} \qquad \text{(S2)}$$

Inspection of Equations S1 and S2 reveals that self-broadening by $H_2$ is less important than foreign-broadening when $H_2$ is the secondary greenhouse gas. This is because the pathlength ($W_{H2}$) is a linear function of the mixing ratio, and $\tau_p$ is a quadratic function of $f_{H2}$. In contrast, $\tau_l$ is approximately linear in $f_{H2}$ at low mixing ratios. Furthermore, a comparison of the tabulated data for self- and foreign-broadening of $H_2$(Tables I and II, respectively) shows that $H_2$ foreign broadening is generally more effective at any given temperature and frequency. We conducted a sensitivity study to quantify the effect of removing the $H_2$ self-broadening component. Our results showed that the maximum surface temperatures obtained in Figure 2 (main text) decreased by no more than 2-3 degrees when this was done.

3) Both Halevy et al.[72] and Wordsworth et al.[64] have pointed out that our old climate model, which was derived from that of Kasting et al.[51], may have significantly overestimated absorption of thermal-IR radiation by CIA bands of $CO_2$. Consequently, we replaced our old $CO_2$ CIA parameterization with the one described in Wordsworth et al.[64]. $CO_2$ CIA becomes significant at higher pressures and consists of two separate effects: 1) close collisions between $CO_2$ molecules that induce temporary dipoles, and 2) colliding molecules that form $CO_2$-$CO_2$ dimers. These two phenomena result in the creation of new absorption bands[73,74,75] . Our revised parameterization (hereafter, GBB after ref. 64) computes the optical depth ($\tau_{CO2}$) from the following formula:

$$\tau_{co_2} = C_i \frac{n}{n_o} W_{co_2} \qquad \text{(S3)}$$

Here $C_i$ is a constant (see below) with units of amagat$^{-2}$ cm$^{-1}$, and $W_{CO2}$ is the $CO_2$ pathlength in atmosphere-centimeters.

The new $CO_2$-$CO_2$ collision-induced absorption coefficients, $C_i$, used in Eq. S3 are tabulated below (Table III). Within the 0-495 cm$^{-1}$ spectral region we replaced the calculated values at 150 K with those at 200 K. We did this because the model of Gruszka and Borysow[73,74] may be unreliable below ~200 K. However, because the pressures associated with such temperatures are very small on Mars, collisions should be very infrequent. Thus, CIA is unimportant at these lower temperatures and our results are insensitive to the absorption coefficient values at T = 150 K.

4) We also incorporated near-IR $CO_2$ CIA from the 1.2, 1.73, and 2.3 micron regions using tabulated values from previous Venusian studies[76-79]. The temperature dependence of near-IR $CO_2$ CIA is difficult to measure experimentally and the values are poorly known. However, as the resultant optical depths were of order $10^{-4}$ to $10^{-3}$, these bands had only a small effect on our simulations.

5) The Shomate Equation (http://www.vscht.cz/fch/cz/pomucky/fchab/Shomate.html) was used to calculate new heat capacity relationships for $CO_2$, $H_2$, and $H_2O$. Notably, at low temperatures, $c_p$ for $CO_2$ decreased by ~30% relative to values in our previous model. This increased the dry adiabatic lapse rate, $g/c_p$, by an equivalent amount but had surprisingly little effect on computed surface temperatures, apparently because the steeper lapse rate in the upper troposphere was largely compensated by a decrease in tropopause height. (As mentioned earlier, our model assumes a moist adiabatic lapse rate, but this relaxes to a dry adiabat in regions where $CO_2$ is not condensing and where $H_2O$ is scarce).

6) Our climate model now includes Rayleigh scattering by $H_2$[80]. This addition had a negligible effect on our results.

7) The Rayleigh scattering coefficient of water vapor was taken from measured values[81-83]. Our previous climate model had used the same value as for air. Again, this had only a small effect on the results, because $H_2O$ is always a minor constituent in these calculations. But the effect of this change on a warm, moist atmosphere would be significant, as the coefficient for $H_2O$ is ~30% lower than that for air.

8) Our new model now incorporates the decrease of gravity with altitude, which tends to slightly decrease thermal infrared emission, because as the gravity decreases, the optical depth at a given pressure level increases. This effect is more pronounced for smaller planets such as Mars[59].

9) Our model can now perform calculations involving multiple solar zenith angles, increasing the accuracy of shortwave absorption. We performed a sensitivity study using 6 gauss points, which slightly decreases the $H_2$ and $pCO_2$ thresholds required to reach the freezing point (Fig. S8).

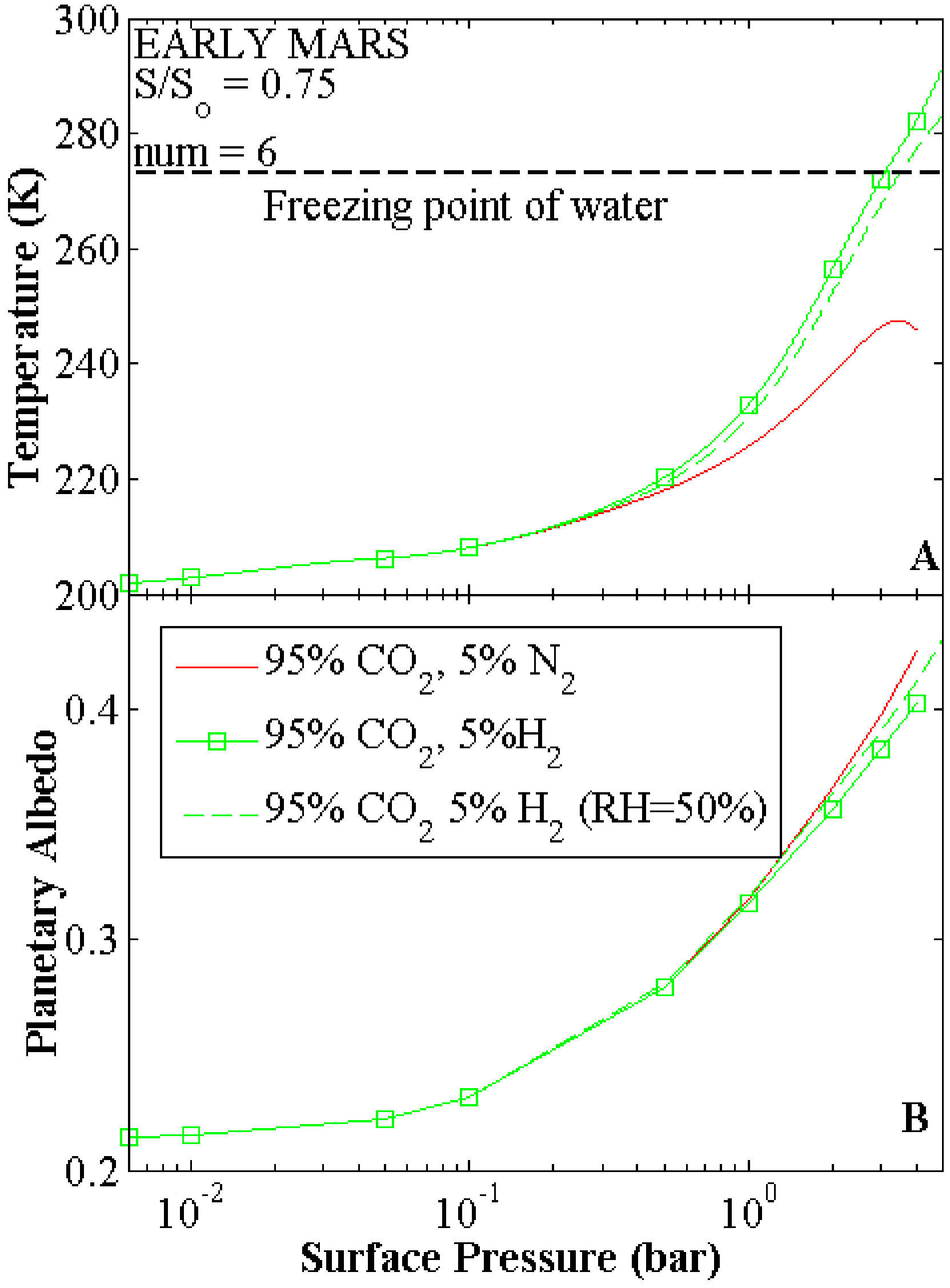

**Figure S8:** Surface temperature and planetary albedo as a function of surface pressure for two 95% $CO_2$, fully-saturated early Mars ($S/S_o = 0.75$) atmospheres containing (a) 5% $N_2$, (b) 5% $H_2$, and (c) 5% $H_2$ with a relative humidity of 50%. The assumed surface albedo is the same as that used in Fig. 2 of the main text. A 6-point gaussian integration scheme was used over the sun-lit hemisphere (num = 6). The cosine-weighted average solar flux factor was ~0.503, which compares to the true value of 0.50. The mean temperature for the baseline 5% $N_2$ case rose from 230 K (Fig. 2 in main text) to 246 K here. With six gauss points, even the 50% relative humidity scenario reaches the freezing point with 5% $H_2$ and just under ~3 bars total surface pressure.

## 2. OUTGASSING OF $H_2$ AND THE ATMOSPHERIC REDOX BUDGET

As discussed in the main text, $H_2$ and other reduced gases should have been outgassed from the reduced martian mantle. The $H_2$:$H_2O$ ratio in these gases would have been determined by reaction (1) in the main text: $2\,H_2O \xleftrightarrow{K_1} 2\,H_2 + O_2$, for which the equilibrium constant $K_1$ is given by

$$K_1 = \frac{p_{H_2}^{\ 2} \cdot fO_2}{p_{H_2O}^{\ 2}} = e^{-\Delta G_1^0/(RT)} \qquad \text{(S4)}$$

Here, $\Delta G_1^0$ is the change in Gibbs free energy for the reaction at standard state. The free energies of formation of $H_2$ and $O_2$ are defined as zero at all temperatures. Using data from the NIST-JANAF thermochemical tables[84], the free energy of formation of $H_2O$ at 1200°C is $\Delta G_f^0(H_2O) =$ −165.58 kJ/mol, so the free energy change for the reaction is just

$$\Delta G_1^0 = -2 \cdot \Delta G_f^0(H_2O) = 331.16 \text{ kJ/mol} \qquad \text{(S5)}$$

Using Eq. S4 above, the equilibrium constant at 1200°C is $K_1 \cong 1.80\times10^{-12}$ atm. This value of the equilibrium constant is used in the main text.

So far, we have only accounted for $H_2$ itself. But, other reduced gases also enter into the atmospheric redox budget, because they can be oxidized by the byproducts of $H_2O$ and $CO_2$ photolysis to produce $H_2$. By adopting a methodology in which certain gases ($H_2O$, $CO_2$, $N_2$, and $SO_2$) are assigned "neutral" oxidation states, one can evaluate the contribution of other gases to the atmospheric redox budget[85,86]. For example, $H_2S$ can be converted to $SO_2$ and $H_2$ by the reaction sequence (see reaction rates in ref. 87)

$H_2S + h\nu \rightarrow HS + H$

$H_2O + h\nu \rightarrow H + OH$ (×2)

$CO_2 + h\nu \rightarrow CO + O$

$$HS + O \rightarrow H + SO$$

$$SO + OH \rightarrow SO_2 + H$$

$$\mathbf{H + CO + M \rightarrow HCO + M} \qquad \mathbf{(\times 3)}$$

$$\mathbf{H + HCO \rightarrow H_2 + CO} \qquad \mathbf{(\times 3)}$$

$$\underline{CO + OH \rightarrow CO_2 + H}$$

$$\underline{\text{Net}}\text{:}\; H_2S + 2\,H_2O \rightarrow SO_2 + 3\,H_2 \qquad \text{(S6)}$$

This net reaction is the same one quoted in the main text. The particular reaction sequence shown here is by no means the only one that would have operated in the primitive atmosphere, but it is a particularly effective one because the two boldface reactions provide a catalytic cycle for producing $H_2$ (ref. 88). $H_2S$ can also react directly with OH to produce $H_2O$ + HS, and H can react with HS to produce $H_2$ + S. Regardless of how the actual reaction path proceeds, each mole of $H_2S$ outgassed is equivalent to 3 moles of outgassed $H_2$, if the sulfur is removed as $SO_2$. If the $H_2S$ is oxidized to sulfate, then an additional mole of $H_2$ is produced. Similarly, for $CH_4$ and CO, we can write

$$CH_4 + 2\,H_2O \rightarrow CO_2 + 4\,H_2 \qquad \text{(S7)}$$

$$CO + H_2O \rightarrow CO_2 + H_2 \qquad \text{(S8)}$$

Thus, one mole of $CH_4$ yields 4 moles of $H_2$, and one mole of CO yields one mole of $H_2$.

All three of these additional reduced gases would have helped to build up the inventory of atmospheric $H_2$. In the main text, we estimated the rate of $H_2$ outgassing on early Mars by making an analogy to modern Earth, and we assumed that outgassing of $H_2S$ scaled proportionately with $H_2$. For the carbon-bearing gases, we can make a similar analogy to modern Earth. On Earth, the main carbon-bearing volcanic gas is $CO_2$, for which the estimated global release rate is $(7.5 \pm 2) \times 10^{12}$ mol/yr [89], or ~$2.8 \times 10^{10}$ $cm^{-2}s^{-1}$. (We will use areal outgassing rates here, so as not to have to account for the different surface areas of Earth and Mars.) On early Mars, because of the planet's reduced mantle, carbon would have been outgassed as a combination of CO and $CH_4$[90]. Wetzel et al. estimate that initial crustal formation would have released 1 bar of CO and 1.3 bar of $CH_4$. By converting from partial pressure to volume units (thereby accounting for the mass of O and H), we calculate that 30% of the outgassed carbon would have been released as CO, and 70% would have been released as $CH_4$. If we assume that early Mars outgassed carbon at the same rate per unit area as modern Earth—the same assumption made in the main text for total hydrogen--the equivalent $H_2$ outgassing rate can be estimated by multiplying the $CO_2$ outgassing rate given above by 0.3 for CO and by 0.7×4 for $CH_4$, where the factor of 4 comes from the stoichiometry of reaction S7. The net equivalent outgassing of $H_2$ is then $8.7\times10^{10}$ $cm^{-2}s^{-1}$. By comparison, the estimates given for combined ($H_2$ + $H_2S$) outgassing in the main text were $1\times10^{10}$ $cm^{-2}$ $s^{-1}$ for modern Earth, $2\times10^{11}$ $cm^{-2}$ $s^{-1}$ for early Mars at a mantle $fO_2$ of IW+1, and $4\times10^{11}$ $cm^{-2}$ $s^{-1}$ for early Mars at IW−1. So, outgassing of $CH_4$ and CO would have made an appreciable, but not dominant, contribution to the atmospheric redox budget of early Mars.

## 3. EFFECT OF $CH_4$ ON CLIMATE IN A DOMINANTLY $CO_2$-$H_2$ EARLY MARS ATMOSPHERE

As discussed above, if Mars' early mantle was highly reduced, some C would have been outgassed in the form of $CH_4$. Thus, $CH_4$ would be expected to have contributed to the greenhouse effect. Here, we assess the climate effect of adding 1% $CH_4$ and various combinations of $CO_2$ and $H_2$ to representative fully-saturated 3-bar atmospheres with a surface temperature of 230 K. The instantaneous radiative forcing, defined here as the reduction in outgoing infrared radiation at the top of the atmosphere, caused by such an addition is shown in Table IV. Any remaining gas in these calculations is assumed to be $N_2$. In all cases, the additional radiative forcing from $CH_4$ is no more than 1 W/m$^2$. By comparison, $H_2$ absorption produced 6 - 22 W/m$^2$ forcing for $H_2$ concentrations of 5-20% (Fig. S6).

Selected temperature-altitude profiles for the 95% $CO_2$, 5% $H_2$ case are given in Fig. S9. $H_2O$ and $CO_2$ cloud-forming regions are shown for the 3-bar case. Clouds are not included explicitly in the radiative transfer model, but their formation does affect the tropospheric lapse rate. As noted above, the greenhouse contribution from $CH_4$ is expected to be small in a dense $CO_2$ early Mars atmosphere. However, $CH_4$ absorbs strongly in the near-IR and could have contributed significantly to heating of the mid and upper atmosphere, thereby inhibiting $CO_2$ cloud formation (Fig. S10). As previous studies[56,57,91] have suggested that backscattering from $CO_2$ ice clouds would warm the surface under high fractional cloud cover ($\geq 0.5$), this could conceivably result in surface cooling as compared to the zero-$CH_4$ case, but perhaps not by that much, as a recent study[92] suggests that the two-stream method may have overestimated the warming coming from the backscatter of $CO_2$ ice crystals. To determine how important this process might be, this effect should be investigated with a more elaborate 3-D climate model that includes clouds, such as the model described in refs. 13 and 14 in the main text.

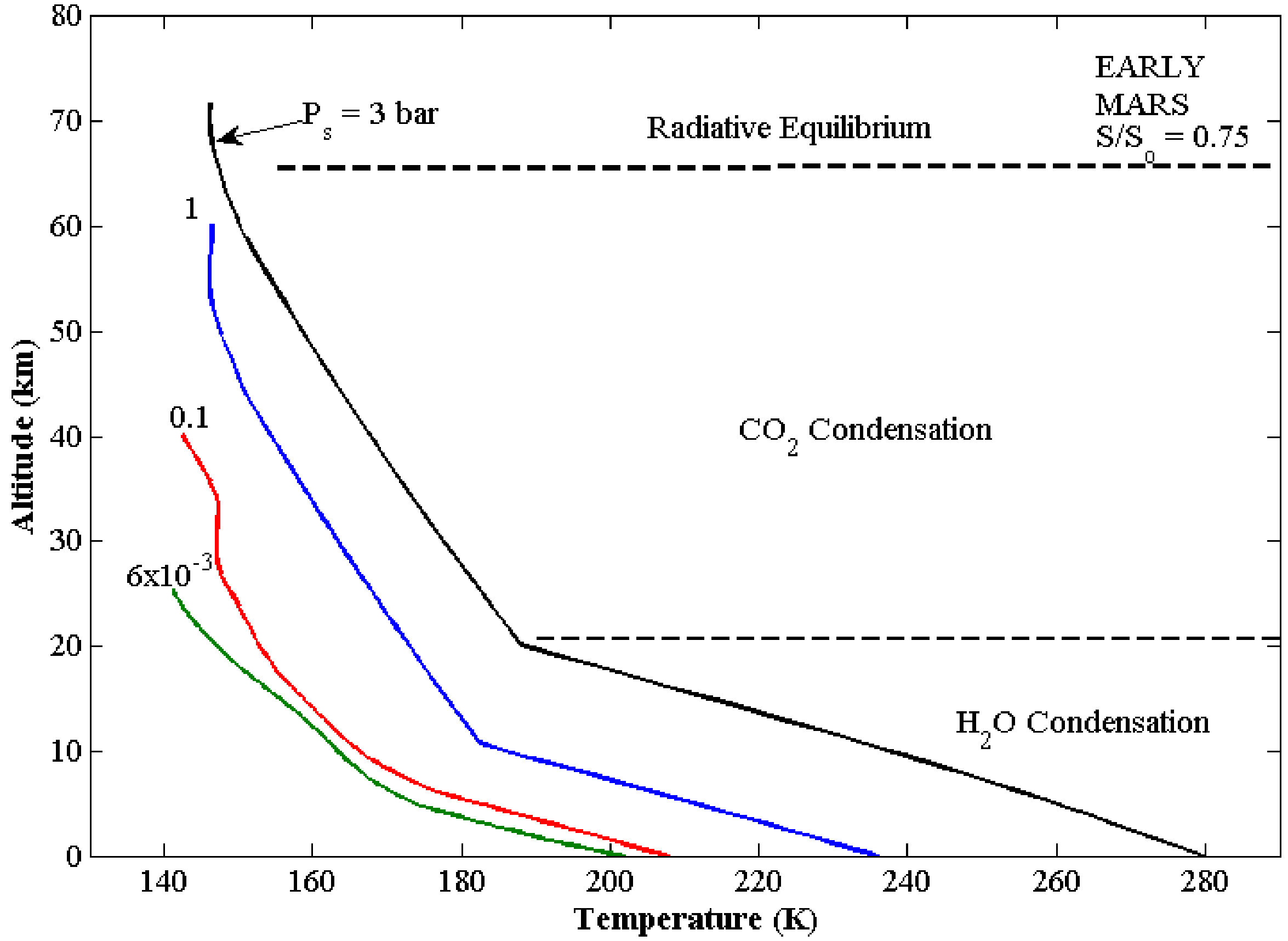

**Figure S9:** Selected vertical temperature profiles for the fully-saturated 95% $CO_2$ 5% $H_2$ case shown in Fig. 2 of the main text. As the total atmospheric pressure increases from $6\times10^{-3}$ to 3 bar, the magnitude of the greenhouse effect from the combined effects of the $CO_2$ greenhouse and $H_2$ collision-induced absorption, is ~70 degrees.

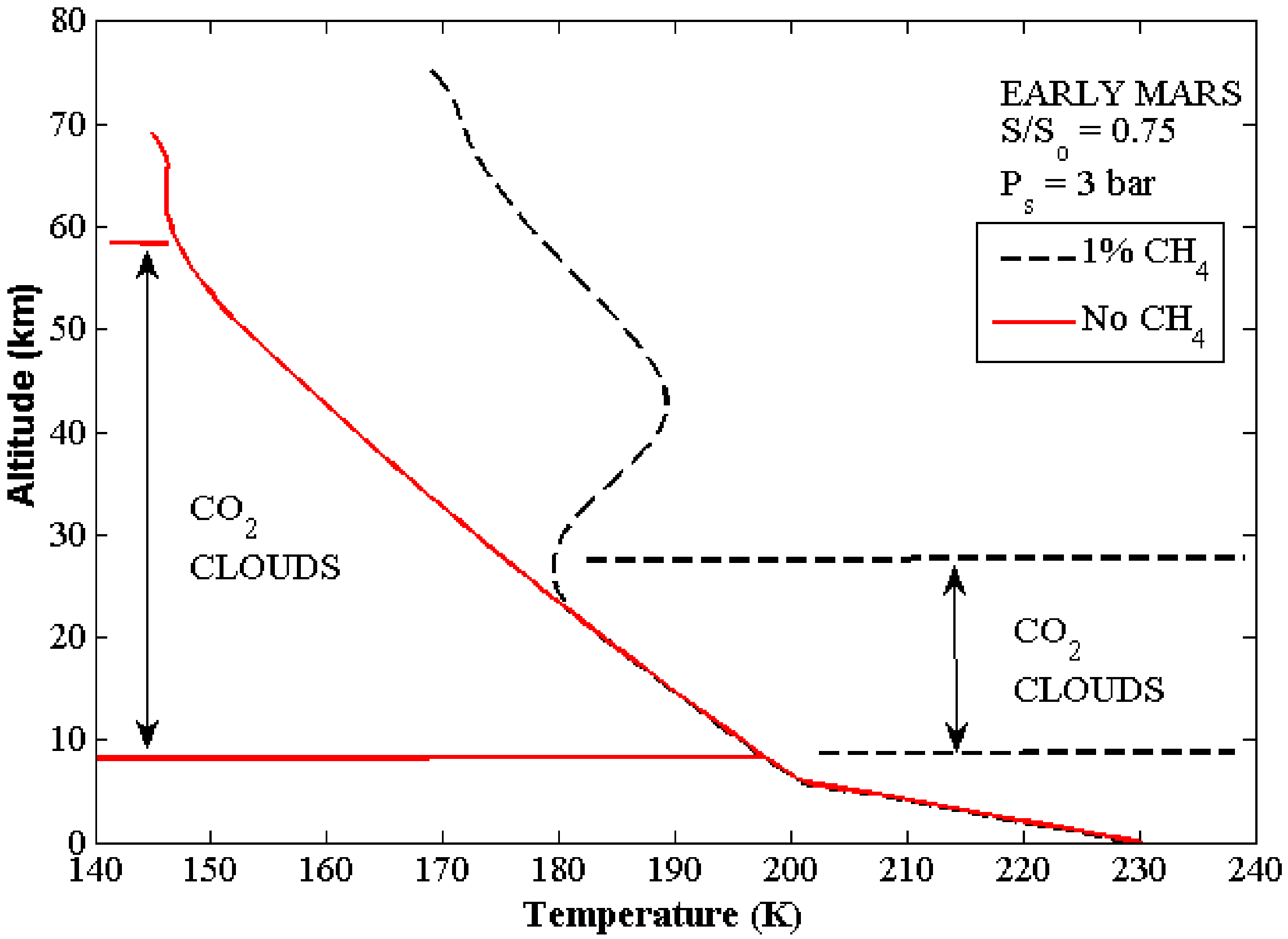

**Figure S10:** Temperature-altitude profiles for fully-saturated 3-bar $CO_2$ early Mars (S/So = 0.75) atmospheres with (a) 1% $CH_4$ and (b) no $CH_4$. The surface albedo is the same as in Fig. 2 in the main text. The calculated surface temperature is ~230 K in both cases. Addition of 1% $CH_4$ has little direct effect on surface temperature, but it decreases the size of the $CO_2$ cloud deck from ~50 km to ~20 km. This effect may prove important in 3D GCMs that include clouds and in assessing the intensity of $CO_2$ condensation at the poles and at higher elevations. Sensitivity studies that included both self-broadened and foreign-broadened $CH_4$ CIA (assuming $CO_2$-$CH_4$ broadens as well as $N_2$-$CH_4$) had a small effect, increasing surface temperatures to just over 231 K. The $CH_4$ self-broadening component is small because $CH_4$ composes only 1% of the atmosphere. $N_2$-$CH_4$ CIA band absorbs most strongly in the far-IR, which is masked by the pure rotation band of water.

## 4. TABLES OF CIA DATA

**Table IA:** $N_2$-$H_2$ Collision-induced absorption coefficients in the 0 - 1000 $cm^{-1}$ spectral range.

| | **T = 150K** | **T = 200K** | **T = 300K** |
|---|---|---|---|
| **Wavenumber interval ($cm^{-1}$)** | **CIA ($amagat^{-2}cm^{-1}$)** | **CIA ($amagat^{-2}cm^{-1}$)** | **CIA ($amagat^{-2}cm^{-1}$)** |
| 0 - 40 | $6.32x10^{-7}$ | $4.33x10^{-7}$ | $2.54x10^{-7}$ |
| 40 - 100 | $2.63x10^{-6}$ | $2.03x10^{-6}$ | $1.35x10^{-6}$ |
| 100 - 160 | $3.08x10^{-6}$ | $2.92x10^{-6}$ | $2.45x10^{-6}$ |
| 160 - 220 | $2.00x10^{-6}$ | $2.24x10^{-6}$ | $2.41x10^{-6}$ |
| 220 - 280 | $1.73x10^{-6}$ | $1.94x10^{-6}$ | $2.17x10^{-6}$ |
| 280 - 330 | $4.26x10^{-6}$ | $3.58x10^{-6}$ | $2.89x10^{-6}$ |
| 330 - 380 | $1.13x10^{-5}$ | $7.42x10^{-6}$ | $4.45x10^{-6}$ |
| 380 - 440 | $7.13x10^{-6}$ | $5.50x10^{-6}$ | $4.07x10^{-6}$ |
| 440 - 495 | $3.93x10^{-6}$ | $4.19x10^{-6}$ | $4.36x10^{-6}$ |
| 495 - 545 | $7.84x10^{-6}$ | $8.77x10^{-6}$ | $8.60x10^{-6}$ |
| 545 - 617 | $2.78x10^{-5}$ | $2.46x10^{-5}$ | $1.85x10^{-5}$ |
| 617 - 667 | $1.87x10^{-5}$ | $1.82x10^{-5}$ | $1.54x10^{-5}$ |
| 667 - 720 | $7.47x10^{-6}$ | $8.44x10^{-6}$ | $8.84x10^{-6}$ |
| 720 - 800 | $3.24x10^{-6}$ | $4.47x10^{-6}$ | $5.91x10^{-6}$ |
| 800 - 875 | $2.72x10^{-6}$ | $4.07x10^{-6}$ | $5.62x10^{-6}$ |
| 875 - 940 | $9.67x10^{-7}$ | $1.73x10^{-6}$ | $3.22x10^{-6}$ |
| 940 - 1000 | $4.60x10^{-7}$ | $1.10x10^{-6}$ | $3.03x10^{-6}$ |

**Table IB:** $N_2$-$H_2$ Collision-induced absorption coefficients in the 1000 - 2050 $cm^{-1}$ spectral range.

| | **T = 150K** | **T = 200K** | **T = 300K** |
|---|---|---|---|
| **Wavenumber interval ($cm^{-1}$)** | **CIA ($amagat^{-2}cm^{-1}$)** | **CIA ($amagat^{-2}cm^{-1}$)** | **CIA ($amagat^{-2}cm$-$^{1}$)** |
| 1000 - 1065 | $5.07x10^{-7}$ | $1.53x10^{-6}$ | $4.45x10^{-6}$ |
| 1065 - 1108 | $3.05x10^{-7}$ | $1.02x10^{-6}$ | $3.33x10^{-6}$ |
| 1108 - 1200 | $1.19x10^{-7}$ | $3.99x10^{-7}$ | $1.55x10^{-6}$ |
| 1200 - 1275 | $4.43x10^{-8}$ | $1.43x10^{-7}$ | $7.03x10^{-7}$ |
| 1275 - 1350 | $2.03x10^{-8}$ | $6.40x10^{-8}$ | $3.56x10^{-7}$ |
| 1350 - 1450 | $9.02x10^{-9}$ | $2.59x10^{-8}$ | $1.74x10^{-7}$ |
| 1450 - 1550 | $6.50x10^{-9}$ | $1.38x10^{-8}$ | $9.44x10^{-8}$ |
| 1550 - 1650 | $1.63x10^{-8}$ | $1.81x10^{-8}$ | $4.35x10^{-8}$ |
| 1650 - 1750 | $1.45x10^{-8}$ | $1.53x10^{-8}$ | $2.53x10^{-8}$ |
| 1750 - 1850 | $5.69x10^{-9}$ | $7.19x10^{-9}$ | $1.26x10^{-8}$ |
| 1850 - 1950 | $2.15x10^{-8}$ | $3.18x10^{-9}$ | $6.29x10^{-9}$ |
| 1950 - 2050 | $1.40x10^{-9}$ | $2.3x10^{-9}$ | $4.49x10^{-9}$ |

**Table IIA:** $H_2$-$H_2$ Collision-induced absorption values in the 0- 1000 $cm^{-1}$ spectral range.

| | **T = 150K** | **T = 200K** | **T = 300K** |
|---|---|---|---|
| **Wavenumber interval ($cm^{-1}$)** | **CIA ($amagat^{-2}cm^{-1}$)** | **CIA ($amagat^{-2}cm^{-1}$)** | **CIA ($amagat^{-2}cm^{-1}$)** |
| 0 - 40 | $4.16x10^{-8}$ | $3.21x10^{-8}$ | $2.16x10^{-8}$ |
| 40 - 100 | $1.67x10^{-7}$ | $1.48x10^{-7}$ | $1.16x10^{-7}$ |
| 100 - 160 | $2.39x10^{-7}$ | $2.46x10^{-7}$ | $2.34x10^{-7}$ |
| 160 - 220 | $2.60x10^{-7}$ | $2.97x10^{-7}$ | $3.23x10^{-7}$ |
| 220 - 280 | $4.29x10^{-7}$ | $4.59x10^{-7}$ | $4.73x10^{-7}$ |
| 280 - 330 | $1.12x10^{-6}$ | $9.48x10^{-7}$ | $7.82x10^{-7}$ |
| 330 - 380 | $2.22x10^{-6}$ | $1.64x10^{-6}$ | $1.19x10^{-6}$ |
| 380 - 440 | $2.20x10^{-6}$ | $1.81x10^{-6}$ | $1.51x10^{-6}$ |
| 440 - 495 | $1.90x10^{-6}$ | $1.62x10^{-6}$ | $1.51x10^{-6}$ |
| 495 - 545 | $2.93x10^{-6}$ | $3.91x10^{-6}$ | $3.15x10^{-6}$ |
| 545 - 617 | $6.14x10^{-6}$ | $5.84x10^{-6}$ | $4.97x10^{-6}$ |
| 617 - 667 | $5.77x10^{-6}$ | $5.71x10^{-6}$ | $5.11x10^{-6}$ |
| 667 - 720 | $3.61x10^{-6}$ | $3.97x10^{-6}$ | $4.13x10^{-6}$ |
| 720 - 800 | $2.11x10^{-6}$ | $2.63x10^{-6}$ | $3.24x10^{-6}$ |
| 800 - 875 | $1.40x10^{-6}$ | $1.93x10^{-6}$ | $2.68x10^{-6}$ |
| 875 - 940 | $8.83x10^{-7}$ | $1.30x10^{-6}$ | $2.10x10^{-6}$ |
| 940 - 1000 | $6.15x10^{-7}$ | $9.58x10^{-7}$ | $1.85x10^{-6}$ |

**Table IIB:** $H_2$-$H_2$ Collision-induced absorption values in the 1000 - 2050 $cm^{-1}$ spectral range.

| | **T = 150K** | **T = 200K** | **T = 300K** |
|---|---|---|---|
| **Wavenumber interval ($cm^{-1}$)** | **CIA ($amagat^{-2}cm^{-1}$)** | **CIA ($amagat^{-2}cm^{-1}$)** | **CIA ($amagat^{-2}cm^{-1}$)** |
| 1000 - 1065 | $4.39x10^{-7}$ | $8.08x10^{-7}$ | $1.84x10^{-6}$ |
| 1065 - 1108 | $3.28x10^{-7}$ | $6.47x10^{-7}$ | $1.59x10^{-6}$ |
| 1108 - 1200 | $2.79x10^{-7}$ | $4.76x10^{-7}$ | $1.14x10^{-6}$ |
| 1200 - 1275 | $1.93x10^{-7}$ | $3.03x10^{-7}$ | $7.37x10^{-7}$ |
| 1275 - 1350 | $1.06x10^{-7}$ | $1.82x10^{-7}$ | $4.91x10^{-7}$ |
| 1350 - 1450 | $6.65x10^{-8}$ | $1.19x10^{-7}$ | $3.26x10^{-7}$ |
| 1450 - 1550 | $3.70x10^{-8}$ | $7.00x10^{-8}$ | $2.05x10^{-7}$ |
| 1550 - 1650 | $2.41x10^{-8}$ | $4.67x10^{-8}$ | $1.38x10^{-7}$ |
| 1650 - 1750 | $1.71x10^{-8}$ | $3.17x10^{-8}$ | $9.19x10^{-8}$ |
| 1750 - 1850 | $1.05x10^{-8}$ | $1.90x10^{-8}$ | $5.74x10^{-8}$ |
| 1850 - 1950 | $6.34x10^{-9}$ | $1.18x10^{-8}$ | $3.74x10^{-8}$ |
| 1950 - 2050 | $3.93x10^{-9}$ | $1.37x10^{-8}$ | $2.42x10^{-8}$ |

**Table III:** $CO_2$-$CO_2$ Collision-induced absorption values in the 0- 495cm$^{-1}$ and 1108-1850 cm$^{-1}$ spectral ranges.

| | **T = 150K** | **T = 200K** | **T = 300K** |
|---|---|---|---|
| **Wavenumber interval(cm$^{-1}$)** | **CIA (amagat$^{-2}$cm$^{-1}$)** | **CIA (amagat$^{-2}$cm$^{-1}$)** | **CIA (amagat$^{-2}$cm$^{-1}$)** |
| 0 - 40 | $4.16x10^{-5}$ | $4.16x10^{-5}$ | $1.46x10^{-5}$ |
| 40 - 100 | $9.93x10^{-5}$ | $9.93x10^{-5}$ | $5.09x10^{-5}$ |
| 100 - 160 | $2.18x10^{-5}$ | $2.18x10^{-5}$ | $1.94x10^{-5}$ |
| 160 - 220 | $3.23x10^{-6}$ | $3.23x10^{-6}$ | $5.09x10^{-6}$ |
| 220 -280 | $4.26x10^{-7}$ | $4.26x10^{-7}$ | $1.20x10^{-6}$ |
| 280 - 330 | $6.36x10^{-8}$ | $6.36x10^{-8}$ | $3.06x10^{-7}$ |
| 330 - 380 | $1.1x10^{-8}$ | $1.1x10^{-8}$ | $8.63x10^{-8}$ |
| 380 - 440 | $1.57x10^{-9}$ | $1.57x10^{-9}$ | $2.11x10^{-8}$ |
| 440 - 495 | $2.04x10^{-10}$ | $2.04x10^{-10}$ | $4.80x10^{-9}$ |
| 1108 - 1200 | $3.14x10^{-7}$ | $2.94x10^{-7}$ | $2.93x10^{-7}$ |
| 1200 - 1275 | $7.91x10^{-6}$ | $6.48x10^{-6}$ | $5.8\text{-}x10^{-6}$ |
| 1275 - 1350 | $4.73x10^{-5}$ | $3.55x10^{-5}$ | $2.08x10^{-5}$ |
| 1350 - 1450 | $4.68x10^{-5}$ | $3.52x10^{-5}$ | $2.08x10^{-5}$ |
| 1450 - 1550 | $1.87x10^{-6}$ | $1.62x10^{-6}$ | $1.58x10^{-6}$ |
| 1550 - 1650 | $6.04x10^{-8}$ | $6.04x10^{-8}$ | $6.04x10^{-8}$ |
| 1650 - 1750 | $2.30x10^{-9}$ | $2.30x10^{-9}$ | $2.30x10^{-9}$ |
| 1750 - 1850 | $8.66x10^{-11}$ | $8.66x10^{-11}$ | $8.66x10^{-11}$ |

**Table IV:** $CH_4$ radiative forcing for select atmospheres

| $fCO_2$ | $fH_2$ | $CH_4$ radiative forcing (W/m$^2$) |
|---|---|---|
| 0.95 | 0 | 0.9 |
| 0.94 | .05 | 1 |
| 0.79 | 0.20 | 0.43 |